\documentclass[aps,pra,twocolumn,amsmath,amssymb,footinbib,showpacs]{revtex4-1}

\newcommand{\Jnature}{Nature (London)}
\newcommand{\Jnatphys}{Nature Phys.}

\newcommand{\Jscience}{Science}

\newcommand{\Jprl}{Phys. Rev. Lett.}
\newcommand{\Jpr}{Phys. Rev.}
\newcommand{\Jpra}{Phys. Rev. A}
\newcommand{\Jprb}{Phys. Rev. B}

\newcommand{\Jrmp}{Rev. Mod. Phys.}

\newcommand{\Jnjp}{New J. Phys.}

\newcommand{\Jjetp}{Sov. Phys. JETP}

\newcommand{\Jphystoday}{Phys. Today}

\newcommand{\Jadvatmoloptphys}{Adv. At. Mol. Opt. Phys.}

\newcommand{\JZphysB}{Z. Phys. B}
\usepackage[english]{babel}
\usepackage{latexsym}
\usepackage{graphics}
\usepackage{subfigure}
\usepackage{epsfig}
\usepackage{color}
\usepackage{hyperref}
\hypersetup{
colorlinks=true,
citecolor=blue,
linkcolor=red,
urlcolor=black
}

\renewcommand{\section}[1]{}

\newcommand{\epsfboxmod}[1]{\epsfbox{#1}}

\newcommand{\infig}[2]{\begin{center}
                                    \mbox{ \epsfxsize #1 \epsfboxmod{#2}}
                                      \vspace{-0.8cm}
                                    \end{center}}

\newcommand{\infigbis}[2]{          \mbox{ \epsfxsize #1 \epsfboxmod{#2}}
                                      \vspace{-0.8cm}}

\newcommand{\ie}{{i.e.}}
\newcommand{\eg}{{e.g.}}

\newcommand{\old}[1]{{}}

\renewcommand{\section}[1]{}
\renewcommand{\subsection}[1]{}

\newcommand{\ch}{\textrm{cosh}}
\newcommand{\sh}{\textrm{sinh}}
\newcommand{\e}{\textrm{e}}

\newcommand{\av}[1]{\overline{#1}}

\newcommand{\Vopt}{V}
\newcommand{\Vr}{V_\textrm{\tiny R}}
\newcommand{\sigmar}{\sigma_\textrm{\tiny R}}

\newcommand{\lyap}{\gamma}
\newcommand{\loc}{L_{\textrm{\tiny loc}}}
\newcommand{\lyapE}[1]{\lyap{(#1)}}
\newcommand{\kE}{k_\textrm{\tiny \textit E}}

\newcommand{\lB}{l_\textrm{\tiny B}}
\newcommand{\DB}{D_\textrm{\tiny B}}

\newcommand{\TFCor}[1]{\tilde{C_{#1}}}  %{\hat{C_{#1}}}
\newcommand{\TFcor}[1]{\tilde{c_{#1}}}  %{\hat{c_{#1}}}

\newcommand{\Id}{I_{\textrm{\tiny D}}}

\newcommand{\Escreen}{\mathcal{E}}
\newcommand{\Iscreen}{\mathcal{I}}
\newcommand{\lambdalaser}{\lambda_{\textrm{\tiny 0}}}
\newcommand{\lprop}{l}  %{l}

\newcommand{\kappaNot}{\kappa_\textrm{\tiny 0}}

\newcommand{\densstat}{n_\infty}

\newcommand{\distv}{\mathcal{D}}
\newcommand{\distE}{\mathcal{D}_\textrm{\tiny E}}
\newcommand{\distvi}{\distv_{\textrm{p}}}

\newcommand{\Emaxgas}{E_{\textrm{at}}}

\newcommand{\largdist}{p_\textrm{\tiny w}}
\newcommand{\centredist}{p_\textrm{\tiny at}}

\newcommand{\lyapfit}{\gamma_{\textrm{\tiny fit}}}

\newcommand{\vect}[1]{\mathbf{#1}}  %{\boldsymbol{#1}}

\begin{document}

\title{Anderson localization of matter waves in tailored disordered potentials}

\author{Marie Piraud}
\author{Alain Aspect}
\author{Laurent Sanchez-Palencia}
\affiliation{
Laboratoire Charles Fabry,
Institut d'Optique, CNRS, Univ Paris Sud,
2 avenue Augustin Fresnel,
F-91127 Palaiseau cedex, France}

\date{\today}

\begin{abstract}
We show that, in contrast to immediate intuition, Anderson localization of noninteracting particles induced by a disordered potential in free space can increase (\ie\ the localization length can decrease) when the particle energy increases, for appropriately tailored disorder correlations.
We predict the effect in one, two and three dimensions, and propose a simple method to observe it using ultracold atoms placed in optical disorder.
The increase of localization with the particle energy can serve to discriminate quantum versus classical localization.
\end{abstract}


\maketitle

%%%%%%%%%%%%%%%%%%%%%%%%%%%%%%%%%%%%%%%%%%%%%%%%%%%%%%%%%%%%%%%%%%%%%%
The transport properties of a coherent wave in a disordered medium are inherently determined by interference of multiple scattering paths, which can lead to spatial localization and absence of diffusion~\cite{abrahams2010}.
This effect, known as Anderson localization (AL), was first predicted for electrons in disordered crystals~\cite{anderson1958} and then extended to classical waves~\cite{john1984}, which permitted observation of AL in a variety of systems (see Ref.~\cite{lagendijk2009,*aspect2009,*fallani2008,*lsp2010} and references therein).
The most fundamental features of AL are ubiquity and universality~\cite{evers2008}.
For instance, in conventional cases, all states are known to be localized in one (1D) and two (2D) dimensions, while in three dimensions (3D) the spectrum splits into
regions of localized states
and regions of extended states,
separated by so-called mobility edges~\cite{abrahams1979}.
Nevertheless, the observable features of AL strongly depend on the system details.

Consider a wave propagating among randomly distributed point scatterers (point-impurity disorder).
In the absence of interference, the propagation is dominated by normal diffusion.
It devises a diffusive medium characterized by the length scale (transport Boltzmann mean free path) $\lB = v\tau$,
with
  $v=\vert\partial \omega(k)/\partial k\vert$ the wave velocity
  [$\omega(k)$ is the dispersion relation]
and
  $\tau$ the scattering time.
Then, localization arises from the interference of the diffusive paths.
The more the wavelength exceeds the mean free path, the stronger interference affects the transport.
It can thus be inferred that the Lyapunov exponent (inverse localization length),
which characterizes the localization strength, reads
$\lyap = \lB^{-1} F_d(k\lB)$
where the function $F_d$ strongly depends on the spatial dimension $d$
and is a decreasing function of the interference parameter $k\lB$.
For a particle (scalar matter wave) in free space and a weak point-impurity disorder,
  $v \propto k$,
  $\tau$ is proportional to the inverse of the density of states ($\rho \propto k^{d-2}$),
  as given by the Fermi golden rule,
and finally $\lB \propto 1/k^{d-3}$.
Then, for any $d\leq 3$, $\gamma$ is a decreasing function of $k$.
In other words, the localization gets weaker when the particle energy $E=\hbar\omega(k)=\hbar^2k^2/2m$ increases, which conforms to natural intuition.

This decrease of $\lyap(E)$ with $E$ however relies on the microscopic details of the system,
namely on the dispersion relation and on the properties of the scattering time assumed above,
and can be altered in different ways.
For instance, it does not hold for lattice systems, such as electrons in disordered crystals,
because the band structure leads to a \textit{nonmonotonic} behavior of $v$ versus $E$,
which can lead to a function $\lyap (E)$ approximately symmetric with respect to the band center~\cite{thouless1979,kappus1981}.
In other systems, such as light waves in dielectric media,
$\tau$ shows Mie resonances~\cite{akkermans2006},
leading to a strongly nonmonotonic behavior of $\lyap(E)$.
In this work, we discuss a different effect.
We show that the standard behavior of $\lyap(E)$ for particles in free space
can be inverted (\ie\ localization can get stronger with increasing energy)
by tailoring the disorder correlations.
The basic idea behind our work is that for non-point scatterers,
the structure factor $\TFCor{}(k)$ appears in the denominator of $\tau$.
Then, if the disorder has strong spatial frequency components around a particular value $k_0$, the scattering strength may not vary monotonously with $E$ around $E(k_0)$, and $\lyap(E)$ can then increase with $E$.
In contrast to the cases discussed above (lattice electrons and light waves in dielectric media),
this effect is purely due to the disorder correlations.
We first study the $1$D case, which allows for exact calculations of $\lyap(E)$ and for explicit test of an efficient scheme to observe the effect with ultracold atoms.
We then extend our analysis to $2$D and $3$D systems using the self-consistent theory of AL.
We finally discuss how the increase of $\lyap (E)$ with $E$
can serve to discriminate quantum versus classical localization.

%-----------------------------------------%
\begin{figure*}[t!]
\begin{center}
\infigbis{54em}{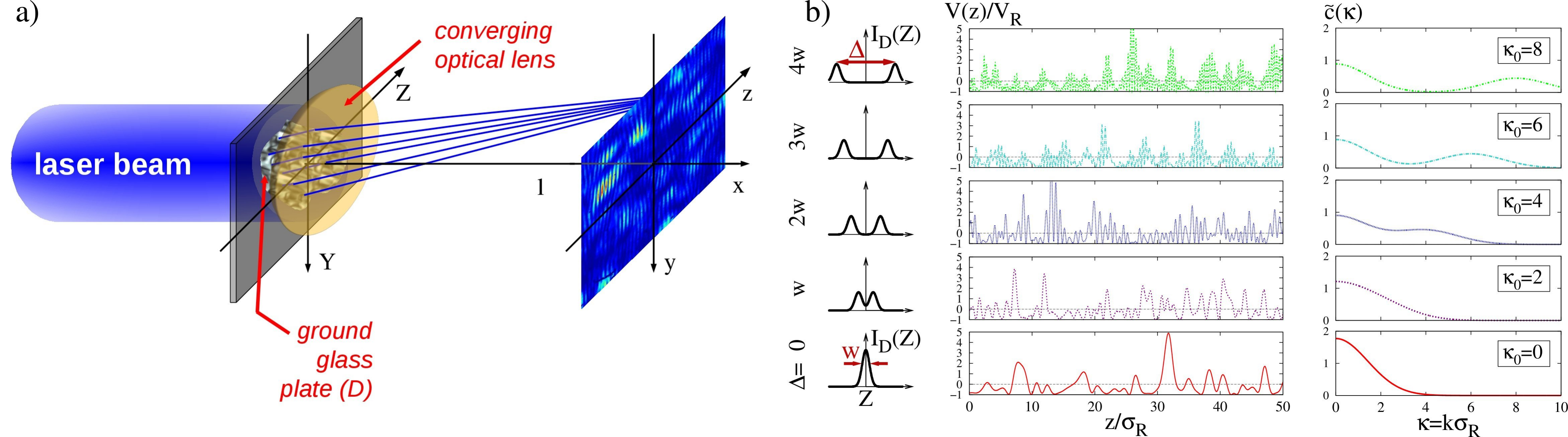}
\end{center}
\vspace{0.1cm}
\caption{
\label{fig:speckle}
(Color online)
Tailoring correlations in speckle potentials.
(a) Optical apparatus:
A laser beam is diffracted by a ground-glass plate diffuser (D)
of pupil function $\Id (\vect{R})$, where $\vect{R} \equiv (Y,Z)$ spans the diffuser,
which imprints a random phase on the various light paths.
The intensity field, $\Iscreen (\vect{r})$, observed
in the focal plane of
a converging lens, is a speckle pattern,
which creates a disordered potential $\Vopt(\vect{r})$ for the atoms.
(b) One-dimensional speckle potentials realized with
a pupil function obtained with two incident Gaussian beams
of waist $w$ and centered at $Z=\pm \Delta/2$.
The figure shows a sketch of $\Id$ ($1^{\textrm{st}}$ column), 
a realization of $\Vopt(z)$ ($2^{\textrm{nd}}$ column),
and the reduced disorder spectrum in $k$-space,
$\TFcor{}(k\sigmar)$ ($3^{\textrm{rd}}$ column)
for various values of $\Delta/w$.
}
\end{figure*}
%-----------------------------------------%

%%%%%%%%%%%%%%%%%%%%%%%%%%%%%%%%%%%%%%%%%%%%%%%%%%%%%%%%%%%%%%%%%%%%%%
A key ingredient of our work is the possibility of tailoring the disorder correlations.
Let us consider speckle potentials~\cite{billy2008,white2009,mrsv2010,kondov2011,jendrzejewski2012}.
A speckle field~\cite{goodman2007} is obtained from a coherent laser beam diffracted by a ground-glass plate
diffuser, which imprints a spatially random phase on the electric field at each point
$\vect{R}$ of its surface [see Fig.~\ref{fig:speckle}(a)].
The diffracted complex electric field $\Escreen(\vect{r})$
at a given observation point $\vect{r}$
is the sum of independent random variables, corresponding to the components
originating from every point $\vect{R}$ of the plate
and interfering in $\vect{r}$.
The atoms are subjected to a potential, which, up to an arbitrary shift, is proportional to the light intensity $\Iscreen \propto |\Escreen|^2$.
We define $\Vopt(\vect{r}) \equiv \Vr \times\{\Iscreen(\vect{r})/\av{\Iscreen}-1\}$,
so that $\av{\Vopt}=0$ and $\av{\Vopt^2}=\Vr^2$
(note that $\av{\Iscreen^2}=2\av{\Iscreen}\, ^2$~\cite{goodman2007}).
The sign of $\Vr$ can be positive or negative depending on the detuning of the laser
with respect to the atomic resonance.
In the paraxial approximation for the scheme of Fig.~\ref{fig:speckle}(a),
the disorder power spectrum (Fourier transform of the disorder correlation function~\footnote{Here, we use $\tilde{f}(\boldsymbol{\kappa})=\int d\vect{u}\, f(\vect{u})\exp (-i\boldsymbol{\kappa}.\vect{u})$.}) in the focal plane $(y,z)$ of the lens is~\cite{goodman2007}
%+++++++++++++++++++++++++++++++++++++++++%
\begin{equation}
\label{eq:c2-autoconv}
\TFCor{}(\vect{k})
\propto
\int d\vect{R} \ \
\Id \big[\vect{R}-({\lambdalaser\lprop}/{4 \pi})\vect{k}\big]
\Id \big[\vect{R}+({\lambdalaser\lprop}/{4 \pi})\vect{k}\big]
\end{equation}
%+++++++++++++++++++++++++++++++++++++++++%
where
$\Id(\vect{R})$ is the pupil function (\ie\ the intensity profile right after the diffusive plate),
$\lambdalaser$ is the laser wavelength, and
$\lprop$ is the focal length.                                                                                      
The major constraints on $\TFCor{}(\vect{k})$ follow from Eq.~(\ref{eq:c2-autoconv})
and from the fact that $\Id (\vect{R})$ is nonnegative and of finite integral.
Firstly, the Cauchy-Schwarz inequality applied to Eq.~(\ref{eq:c2-autoconv})
shows that $\TFCor{}(\vect{k})$
is a decreasing function of $\vert\vect{k}\vert$ for small values of $\vert\vect{k}\vert$.
Secondly, in practice, $\Id (\vect{R})$ decays at long distance
so that $\TFCor{}(\vect{k})$ also
decays in the large $\vert\vect{k}\vert$ limit.
Apart from these constraints, control of $\Id (\vect{R})$ offers freedom for tailoring the disorder power spectrum,
which we will write
$\TFCor{}(\vect{k})=\Vr^2\sigmar^d\,\TFcor{}(\vect{k}\sigmar)$
with $\sigmar$ the correlation length.
We now show that it allows us to strongly affect the qualitative behavior of
AL for noninteracting particles.

\begin{figure}[t!]
\infig{25.5em}{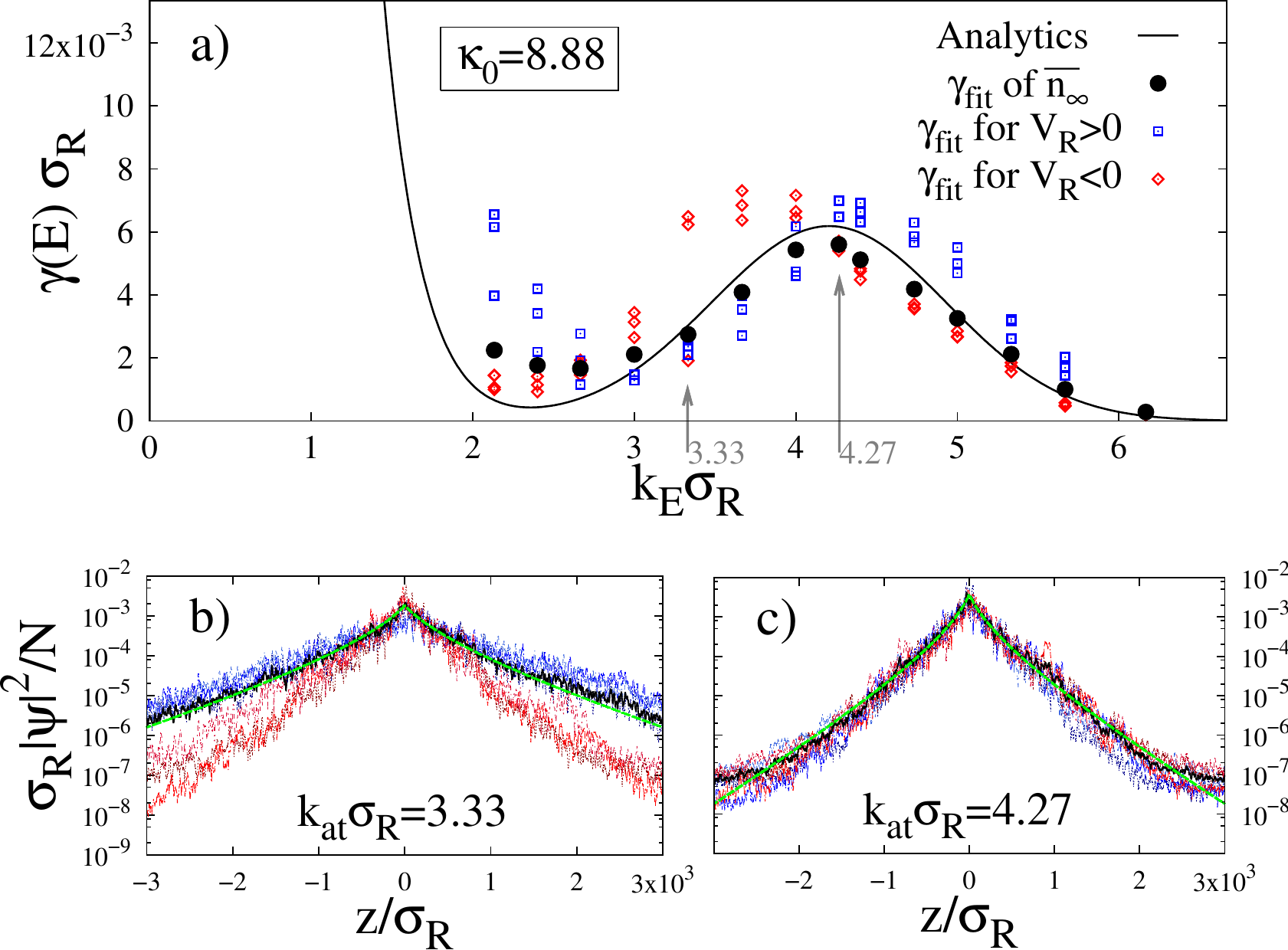}
\vspace{0.2cm}
\caption{
\label{fig:num-stat}
(Color online)
Anderson localization in 1D speckle potentials
with the autocorrelation function~(\ref{eq:f2-gauss}),
$\kappaNot=8.88$ and
$\Vr = \pm 0.72(\hbar^2/m\sigmar^2)$.
(a) Lyapunov exponent versus particle energy as obtained
from Eq.~(\ref{eq:lyap-ordre2}) (solid black line)
and from fits of Eq.~(\ref{eq:gogolin}) to numerical data (points).
(b-c) Stationary density profiles obtained numerically using the
initial state~(\ref{eq:wigngauss})
with $\largdist \sigmar/\hbar=0.24$
and two different values of $\centredist$.
The figures show the results for six realizations of the disorder
[three with $\Vr>0$ (blue upper data) and three with $\Vr<0$ (red lower data)],
the averaged density profile (black central data)
and the fits of $P_\infty(z)$ to the latter (green solid line).
The extracted values of $\lyapfit$ for each realization and for the averaged
profile are reported in (a).
The figure shows a significant increase of the Lyapunov exponent in the energy window
such that $2.3 \lesssim \kE\sigmar \lesssim 4.2$.
}
\end{figure}

%%%%%%%%%%%%%%%%%%%%%%%%%%%%%%%%%%%%%%%%%%%%%%%%%%%%%%%%%%%%%%%%%%%%%%
To start with, consider the 1D case.
Exact calculations can be performed~\cite{lifshits1988} within the Born approximation,
valid for weak disorder [\ie\ for $\lyap (E) \ll \kE, \sigmar^{-1}$ where $\kE=\sqrt{2mE}/\hbar$].
They yield the Lyapunov exponent
%+++++++++++++++++++++++++++++++++++++++++%
\begin{equation}
\lyapE{E} = \loc^{-1}(E) \simeq ({m^2}\Vr^2\sigmar/{2\hbar^4 \kE^2})\TFcor{}(2\kE\sigmar).
\label{eq:lyap-ordre2}
\end{equation}
%+++++++++++++++++++++++++++++++++++++++++%
As can be explicitly seen in Eq.~(\ref{eq:lyap-ordre2}), if the disorder features no particular correlations,
\ie\ if $\TFcor{}(\kappa)$ is a constant or decreasing function of $\kappa$, then $\lyap (E)$ decreases monotonically with $E$,
and the localization is weaker for higher energy, as for point-impurity disorder.
In order to invert this behavior in a given energy window, it is necessary to tailor the disorder correlations so that $\TFcor{}(2\kE\sigmar)$ increases with $\kE$ strongly enough to overcome the $1/\kE^2$ decrease of the prefactor in Eq.~(\ref{eq:lyap-ordre2}).
To do so, we propose to use speckle potentials realized by illuminating the diffusive plate by
two mutually coherent Gaussian laser beams of waist $w$ along $Z$ and centered at $Z=\pm \Delta/2$~\footnote{One can also use a homogeneously illuminated double rectangular aperture
as proposed in Ref.~\cite{plodzien2011} to realize atomic band-pass filters.
We have checked that the two methods lead to qualitatively similar results.
The two-Gaussian scheme allows for more compact formulas and avoids slope breaks
of the $\TFCor{}(k)$ function.}.        
Using Eq.~(\ref{eq:c2-autoconv}), we find
%+++++++++++++++++++++++++++++++++++++++++%
\begin{equation}
\label{eq:f2-gauss}
\TFcor{}(\kappa)=
\frac{\sqrt{\pi}}{4}
\left[ \e^{-\left( \kappa-\kappaNot \right)^2/4} + 2\e^{- \kappa^2/4} + \e^{-\left( \kappa+\kappaNot \right)^2/4} \right]
\end{equation}
%+++++++++++++++++++++++++++++++++++++++++%
with $\sigmar=\lambdalaser\lprop / \pi w$
and
$\kappaNot = 2\Delta/w$, the values of which
can be independently controlled.
The properties of the disordered potentials obtained in this configuration are shown in Fig.~\ref{fig:speckle}(b) for various values of $\kappaNot$.
For $\kappaNot=0$ (lower row),
the disordered potential features structures of typical width $\sigmar$ in real space (central column).
The corresponding power spectrum $\TFCor{}(k)$
has a single Gaussian peak of rms width $\sqrt{2}/\sigmar$ centered in $k=0$ (right column).
For $\kappaNot \neq 0$, the
disordered potential develops additional structures of typical width $\sigmar/\kappaNot$,
corresponding in $\TFCor{}(k)$ to an additional peak centered in $k \simeq \kappaNot/\sigmar$.
For $\kappaNot$ large enough, $\TFCor{}(k)$ shows a clear increase with $k$ in a significant range (upper rows).
For $\kappaNot \gtrsim 5.3$,
we find that it is strong enough that $\lyap (E)$
is nonmonotonic, hence realizing the desired situation
where localization becomes stronger when the particle energy increases.
For instance, for $\kappaNot=8.88$, $\lyap(E)$ shows a significant increase
between $\kE \simeq 2.3\sigmar^{-1}$ and $\kE \simeq 4.2\sigmar^{-1}$
[see Fig.~\ref{fig:num-stat}(a)].

%-----------------------------------------%
%-----------------------------------------%

%%%%%%%%%%%%%%%%%%%%%%%%%%%%%%%%%%%%%%%%%%%%%%%%%%%%%%%%%%%%%%%%%%%%%%
We now discuss how to observe this effect using ultracold gases expanding in the disordered potential.
In the long-time limit, for vanishing interactions and an initial gas of negligible width,
the average spatial density reads~\cite{lsp2007,*lsp2008,piraud2011a}
$n (z,t\rightarrow\infty) = \int dE\ \distE (E) P_\infty(z | E)$,
where $\distE (E)$ is the energy distribution of the atoms and
%+++++++++++++++++++++++++++++++++++++++++%
\begin{eqnarray}
P_\infty(z \vert E)
& = &\frac{\pi^2 \lyap}{8} \int_0^\infty\! \textrm{d}u\ 
u\ \sh (\pi u) \left[ \frac{1+u^2}{1+\ch (\pi u)} \right]^2
\nonumber \\
& & \times \exp\{- (1+u^2) \lyap |z| / 2 \}\,,
\label{eq:gogolin}
\end{eqnarray}
%+++++++++++++++++++++++++++++++++++++++++%
with $\lyap = \lyap(E)$ given by Eq.~(\ref{eq:lyap-ordre2}),
is the probability of quantum diffusion~\cite{gogolin1976a}.
Using the scheme of Ref.~\cite{billy2008},
for which the energy distribution extends from $E=0$ to $E=E_\textrm{max}$,
does not allow us to probe the region where $\lyap (E)$ increases because
the long distance behavior of $n_\infty(z)$
would always be dominated by the energy components with the largest localization lengths,
\ie\ those with the smallest $\lyap (E)$~\cite{lsp2007,*lsp2008,piraud2011a}.
Instead, we propose to use
an atomic energy distribution strongly peaked at a given energy $\Emaxgas$,
so that $n_\infty (z) \simeq P_\infty(z \vert \Emaxgas)$.
It can be realized by
either giving a momentum kick
to a noninteracting initially trapped gas
or using an atom laser,
both with a narrow energy width.
The momentum distribution can be represented by a 1D Gaussian
function of width $\largdist$ centered around
a controllable value $\centredist$~\cite{robins2006,*guerin2006,*bernard2011}:
%+++++++++++++++++++++++++++++++++++++++++%
\begin{equation}
\distvi(p)=({1}/{\sqrt{2\pi} \largdist}) \exp\left[{-( p - \centredist )^2 / 2\largdist^2}\right].
\label{eq:wigngauss}
\end{equation}
%+++++++++++++++++++++++++++++++++++++++++%
For weak disorder, the corresponding energy distribution is weakly affected by the disorder-induced spectral broadening, so that it is strongly peaked at $\Emaxgas \simeq \centredist^2/2m$~\cite{piraud2011a}.

We have performed numerical integration of the time-dependent Schr\"odinger equation
for a particle in the disordered potential with
the initial momentum distribution~(\ref{eq:wigngauss})
and disorder parameters as in Fig.~\ref{fig:num-stat}(a).
During the expansion,
back and forth scattering processes quickly redistribute
left- and right-moving atoms.
The center of the cloud hardly moves
and the wings gradually form a nearly symmetrical
stationary density profile $\densstat(z)$,
shown in Figs.~\ref{fig:num-stat}(b) and (c)
for two values of $\centredist$ and
for six realizations of the disordered potential:
three with blue detuning ($\Vr>0$)
and three with red detuning ($\Vr<0$).
The density profile averaged over the six realizations, $\av{\densstat}(z)$,
is also displayed (black line).
After averaging, we fit $\ln[P_\infty(z)]$
as given by Eq.~(\ref{eq:gogolin})
to $\ln[\av{\densstat}(z)]$
with $\lyap$ as the only fitting parameter.
Although the fits are performed in a limited space window
($-300\sigmar < z < +300\sigmar$,
corresponding to an experimentally accessible width of $1$mm for $\sigmar=1.6\mu$m),
we find that they are good on the total space window ($\vert z \vert$ up to $3000\sigmar$).
As shown in Fig.~\ref{fig:num-stat}(a),
the extracted values $\lyapfit$ (black dots) fairly agree
with Eq.~(\ref{eq:lyap-ordre2}),
except for low energy where the Born approximation breaks down.
The values extracted in the same manner
for each realization of the disordered
potential are also shown (blue squares and red diamonds).
We find nonnegligible difference
between blue and red detunings (see Fig.~\ref{fig:num-stat}),
which can be ascribed to higher-order terms in the Born expansion~\footnote{We estimated the
role of higher order terms in the Born expansion using the approach of Ref.~\cite{lugan2009}.
The calculated corrections are consistent with the numerical results of
Fig.~\ref{fig:num-stat}(a),
including the change of sign found around the local maximum of $\lyap(E)$.}.
Nevertheless, this difference is small and
the strong increase of $\lyap (E)$ appears for each realization
in approximately the same region as predicted by Eq.~(\ref{eq:lyap-ordre2}).
The parameters we used are relevant to current experiments
as regards disorder~\cite{billy2008,mrsv2010}, observable space~\cite{billy2008},
and width of atom lasers~\cite{robins2006,*guerin2006,*bernard2011}.
It validates our proposal to probe the energy dependence of $\lyap(E)$.

%%%%%%%%%%%%%%%%%%%%%%%%%%%%%%%%%%%%%%%%%%%%%%%%%%%%%%%%%%%%%%%%%%%%%%
We now generalize the above results to higher dimensions ($d>1$),
for which the localization scenario is more involved.
At intermediate distance (between the transport mean free path $\lB$
and the localization length $\loc=1/\lyap$),
where interference effects play a negligible role,
the dynamics is diffusive with the diffusion constant
$\DB (E) = (\hbar/m)\kE\lB(E)/d$~\cite{vollhardt1980a,*vollhardt1980b}.
For weak, isotropic disorder
[\ie\ for $\TFcor{}(\vect{\kappa})=\TFcor{}(\vert\vect{\kappa}\vert)$],
one finds~\cite{kuhn2007}
%+++++++++++++++++++++++++++++++++++++++++%
\begin{equation}
\lB^{-1} =
\frac{m^2\Vr^2\sigmar^{d}}{(2\pi)^{d-1} \hbar^4 \kE^{3-d}}
\int d\Omega_d \, (1-\cos\theta) \, \TFcor{} \left(2\kE\sigmar\vert\sin(\theta/2)\vert\right)
\label{eq:lB}
\end{equation}
%+++++++++++++++++++++++++++++++++++++++++%
with $\Omega_d$ the hyperspherical angle in dimension $d$.
On length scales larger than $\lB$, interference effects induce AL,
characterized by the Lyapunov exponent $\lyap$.
The latter can be calculated using the self-consistent theory of AL~\cite{vollhardt1992}.
In 2D one finds, $\lyap (E) = \lB^{-1}\exp (-\pi\kE\lB/2)$.
In 3D, $\lyap (E)$ is the unique solution of
$\left[1  - (\pi/3)(\kE\lB)^2\right] = \lyap\lB\times\arctan (1/\lyap\lB)$,
which exists only below the localization threshold
(mobility edge), \ie\ for $\kE\lB < \sqrt{3/\pi}$~\cite{kuhn2007}.
In both cases we can formally write
$\lyap (E) = \lB^{-1}F_d(\kE\lB)$
with $F_d$ a decreasing function of $\kE \lB$,
consistently with the scaling discussed in the introduction.
It follows from Eq.~(\ref{eq:lB}) that,
if $\TFcor{}(\kappa)$ is as usual a constant or decreasing function of $\kappa$,
then $\lB (E)$ increases with $E$, and
$\lyap (E)$ decreases when $E$ increases.
As for the 1D case, this standard behavior can be changed by tailoring the
disorder correlations so that $\TFCor{}(\vect{k})$ increases strongly enough in a certain window,
and observed in the same way.

\begin{figure}[t!]
\infig{25.5em}{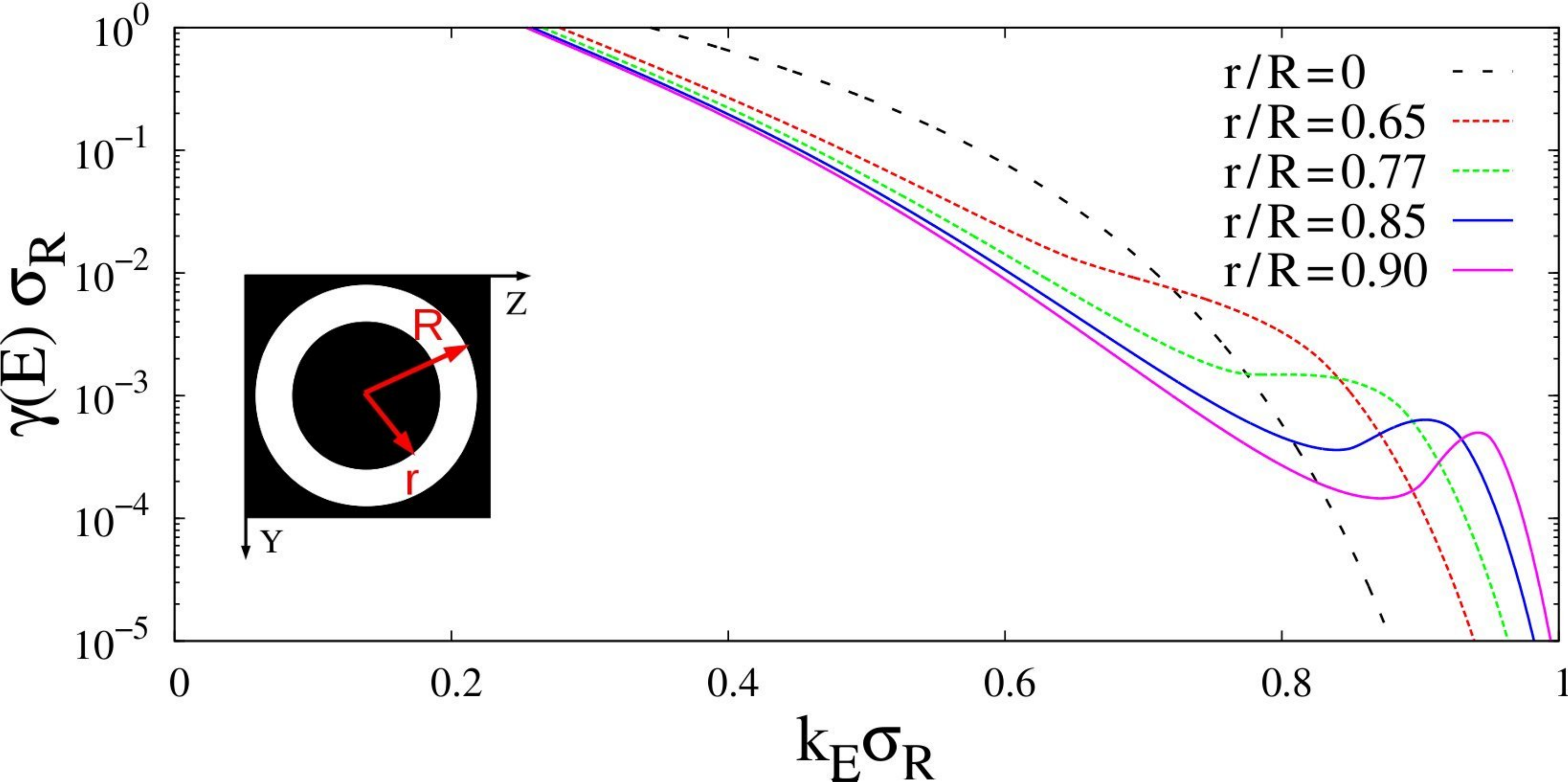}
\vspace{0.2cm}
\caption{
\label{fig:lyap2D}
(Color online)
Lyapunov exponent versus energy
in 2D speckle potentials created with a ring-shaped diffuser
of inner radius $r$
and outer radius $R$
(see Inset), for increasing values of $r/R$ (from the upper line to the lower line on the left-hand-side of the figure)
and $\vert\Vr\vert = 0.25 (\hbar^2/m\sigmar^2)$.
}
\end{figure}

In 2D, we propose to use an isotropic speckle potential created by a uniformly illuminated
ring-shaped diffuser of inner radius $r$ and outer radius $R$
(see Inset of Fig.~\ref{fig:lyap2D}).
For a thin enough ring ($0.77 R \lesssim r < R$),
we find that $\lyap (E)$ is nonmonotonous with a marked local maximum,
so that localization increases with energy in a given window (see Fig.~\ref{fig:lyap2D}).
For the parameters of Fig.~\ref{fig:lyap2D},
$\lyap (E)$ peaks to about $5\times10^{-4}\sigmar^{-1}$,
with $\sigmar = \lambdalaser \lprop/2 \pi R$.
For $\sigmar = 0.25\mu$m,
it corresponds to $\loc \simeq 500\mu$m,
which is within experimental reach~\cite{billy2008}.
Moreover, the width of the maximum is $\Delta k \sim 0.1\sigmar^{-1}$,
which can be probed with the same atom laser as used in Fig.~\ref{fig:num-stat},
the width of which is $\largdist=0.0375\hbar \sigmar^{-1}$
(note that $\sigmar$ is a factor 6.4 larger in the 1D case above).

%-----------------------------------------%
%-----------------------------------------%

\begin{figure}[t!]
\infig{25.5em}{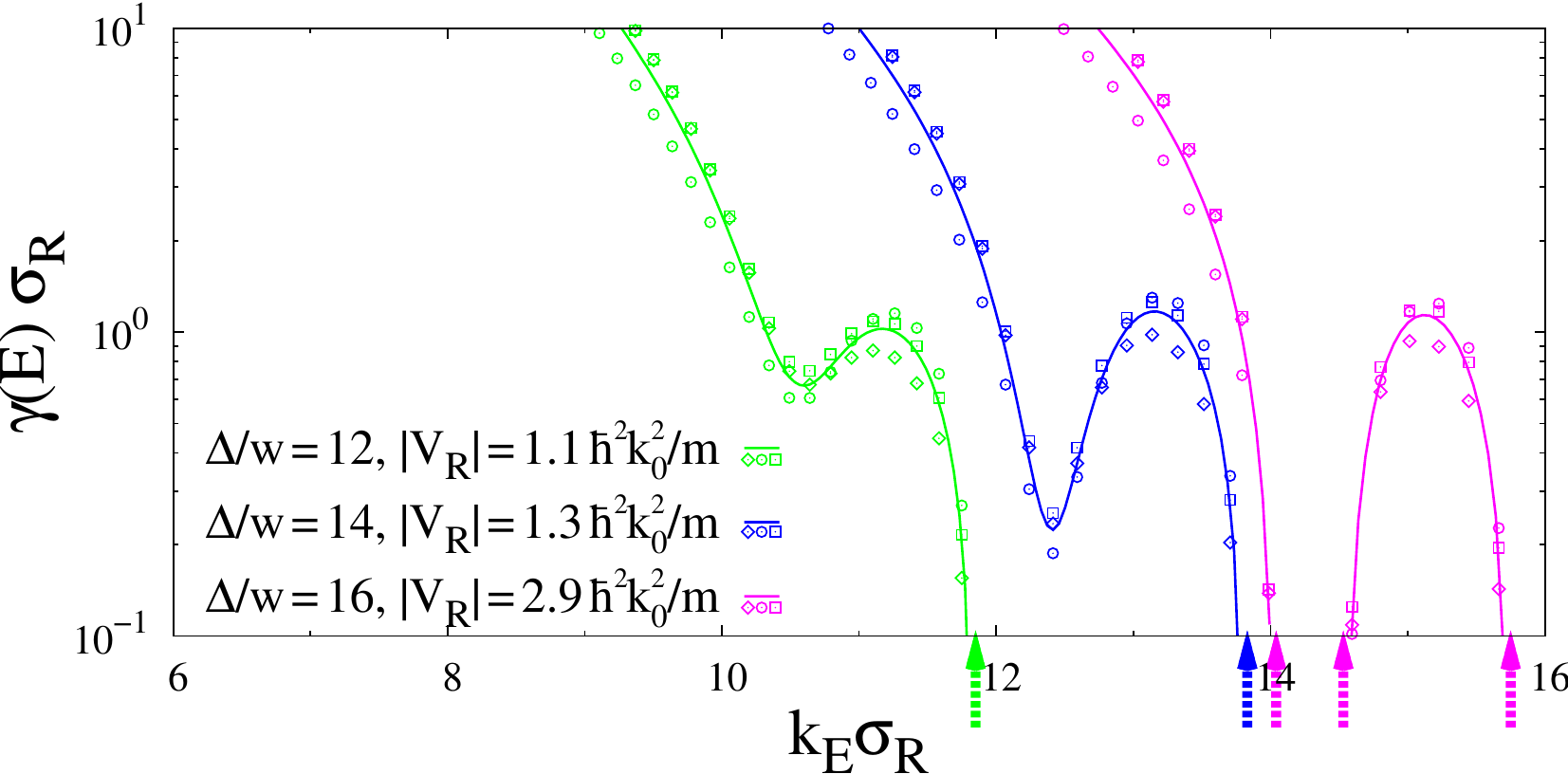}
\vspace{0.2cm}
\caption{
\label{fig:lyap3D}
(Color online)
Lyapunov exponents along the transport eigenaxes ($x$: diamonds; $y$: dots; $z$: squares)
and their geometric average (solid lines)
versus energy in a 3D speckle potential created by two Gaussian beams
(waist $w$ and separation $\Delta$) for various parameters (indicated in the figure).
The arrows indicate the mobility edges.
We recall that $\sigmar=\lambdalaser\lprop / \pi w$.
}
\end{figure}

In 3D, we consider an anisotropic scheme, easier to realize in practice. Then, Eq.~(\ref{eq:lB}) does not hold, but it can be anticipated that tailoring the correlations in anisotropic models of disorder can also invert the standard behavior of $\lyap(E)$.
Consider the 3D speckle potential obtained by one Gaussian laser beam, of waist $w$ (in both transverse directions) [same scheme as in Fig.~\ref{fig:speckle}(a)].
Due to the anisotropy of the disorder, the localization is described by an anisotropic Lyapunov tensor of eigenaxes $x$, $y$ and $z$, and the Lyapunov exponent in all directions increases monotonically with $E$, as shown in Ref.~\cite{piraud2012a}. If we now use two coherent parallel Gaussian beams of same waist $w$ and separated by a distance $\Delta$ along $Z$ (similarly as for the 1D case), the interference between the two speckles create two bumps in $\TFCor{}(\vect{k})$ at $\vect{k} \simeq \pm k_0 \hat{\vect{k}}_z$ with $k_0=\sqrt{2} \pi \Delta/\lambdalaser \lprop$ and $\hat{\vect{k}}_z$ the unit vector in the $k_z$ direction. These two bumps are expected to strongly enhance the localization around the energy $E \propto \hbar^2 k_0^2/2m$.
Figure~\ref{fig:lyap3D} shows the Lyapunov exponents found using
the self-consistent theory of AL for anisotropic disorder~\cite{vollhardt1992,woelfle1984,piraud2012a}.
As expected, the maximum is found for approximately $\kE \propto k_0$.
For all the configurations of Fig.~\ref{fig:lyap3D}, $\lyap (E)$ exhibits, below the right most mobility edge,
a local maximum in each direction, hence realizing the desired effect.
For the parameters of the right (purple) curve, 
$\lyap (E)$ vanishes (\ie\ the localization length diverges) in a given energy window,
thus opening a band of extended states inside the localized region, delimited by two new mobility edges.

%-----------------------------------------%
%-----------------------------------------%

%%%%%%%%%%%%%%%%%%%%%%%%%%%%%%%%%%%%%%%%%%%%%%%%%%%%%%%%%%%%%%%%%%%%%%
In summary, we have shown that the AL of noninteracting quantum particles
(matter waves) induced by a correlated disorder in free space
can increase with the particle energy in a given window.
In contrast to other systems where this behavior is more common,
\eg\ electrons in crystal lattices or light waves in dielectric materials,
it is here purely due to appropriately tailored correlations of the disorder.
We have proposed suitable methods to tailor the correlations in optical disorder,
which require moderate modifications of existing schemes in 
1D~\cite{billy2008},
2D~\cite{mrsv2010},
and 3D~\cite{kondov2011,jendrzejewski2012}.
We have proposed a method to observe it in any dimension, which conversely differs from standard schemes used
so far with ultracold atoms, and explicitly demonstrated its efficiency in the 1D case.

Let us finally discuss how the increase of $\lyap (E)$ with $E$ can serve as a smoking-gun evidence
of quantum localization of particles.
For any experiment on localization, AL should be discriminated from other possible effects. 
For light waves for instance, it is necessary to distinguish it from absorption, which also
produces exponential decay of the intensity. This can be done by analyzing the statistics of
transmission~\cite{chabanov2000,hu2008}.
In contrast, ultracold atoms are not subjected to absorption, but
they can be classically localized (trapped) in potential wells, below the localization threshold.
Then, absence of diffusion and exponential decay of density profiles
can hardly be viewed as indisputable proof of AL.
For instance, classical localization in some
non-percolating media can lead to qualitatively similar effects,
for instance in 2D speckle potentials~\cite{pezze2011b}.
For \textit{any model of disorder} however, the \textit{classical localization length},
defined as the average size of the classically-allowed patches~\cite{zallen1971},
increases with the particle energy.
Hence, the decrease of the \textit{quantum localization length} with the particle energy discussed in this work
has no classical equivalent, and can be viewed as a smoking gun of quantum localization.
This effect could be useful to demonstrate AL, in particular for 2D speckle potentials,
which have a percolation threshold significantly higher than their 3D counterparts.
>From a practical point of view, it does not require accumulation of many statistical data,
in contrast to standard methods used for classical waves~\cite{chabanov2000,hu2008}.

%%%%%%%%%%%%%%%%%%%%%%%%%%%%%%%%%%%%%%%%%%%%%%%%%%%%%%%%%%%%%%%%%%%%%%
We thank P.~Chavel, D.~Cl\'ement and J.-J.~Greffet for enlightening discussions. 
This research was supported by
ERC (contract No.\ 256294),
% the European Research Council (FP7/2007-2013 Grant Agreement No.\ 256294),
ANR (contract No.\ ANR-08-blan-0016-01),
% Agence Nationale de la Recherche (ANR-08-blan-0016-01),
MENRT,
% Minist\`ere de l'Enseignement Sup\'erieur et de la Recherche,
Triangle de la Physique,
and
IFRAF.
% Institut Francilien de Recherche sur les Atomes Froids (IFRAF).
We acknowledge GMPCS computing facilities of the LUMAT federation.
% We acknowledge the use of the computing facility cluster GMPCS of the 
% LUMAT federation (FR LUMAT 2764)

% \bibliography{$HOME/Documents/work/publications/papers/bibliography/biblioLSP}

%

\end{document}